\pdfoutput=1
\documentclass[%
 reprint,
 prd,
 superscriptaddress,
 preprintnumbers,
 bibnotes,
 amsmath,amssymb,
 aps,
]{revtex4-1}
\usepackage{graphicx}
\usepackage{color}
\usepackage{array}
\usepackage{slashed}
\usepackage{tikz}
\usepackage{hyperref}
\usepackage[caption=false]{subfig}
\usepackage{arydshln}
\usepackage{comment}
\usepackage{yhmath}

\usetikzlibrary{shapes.geometric,arrows,decorations.pathmorphing,decorations.markings,patterns,calc}

\usepackage{color}

\newif\ifdraft
\drafttrue

\def\spa#1.#2{\left\langle#1\,#2\right\rangle}
\def\spb#1.#2{\left[#1\,#2\right]}

\newcommand{\eq}{\begin{equation}}
\newcommand{\eqe}{\end{equation}}
\newcommand{\eqa}{\begin{eqnarray}}
\newcommand{\eqae}{\end{eqnarray}}

\newcommand{\bea}{\begin{eqnarray}}
\newcommand{\eea}{\end{eqnarray}}

\newbox\charbox
\newbox\slabox
\def\s#1{{      
        \setbox\charbox=\hbox{$#1$}
        \setbox\slabox=\hbox{$/$}
        \dimen\charbox=\ht\slabox
        \advance\dimen\charbox by -\dp\slabox
        \advance\dimen\charbox by -\ht\charbox
        \advance\dimen\charbox by \dp\charbox
        \divide\dimen\charbox by 2
        \raise-\dimen\charbox\hbox to \wd\charbox{\hss/\hss}
        \llap{$#1$}
}}

\def\be{\begin{equation}}
\def\ee{\end{equation}}
\def\ba{\begin{eqnarray}}
\def\ea{\end{eqnarray}}
\def\nl{\nonumber\\}
\def\ln{\\ \nonumber}

\def\({\left(}
\def\){\right)}
\def\[{\left[}
\def\]{\right]}

\def\<{\langle}
\def\>{\rangle}

\def\bea#1\ea{\begin{eqnarray}#1\end{eqnarray}}
\def\be#1\ee{\begin{equation}#1\end{equation}}
\def\ba#1\ea{\begin{align}#1\end{align}}

\def\nl{\nonumber\\} 
\def\non{\nonumber}

\def\yz#1\yz {{\color{blue} [[YZ: #1]] }}

\usepackage{cleveref}

\usepackage{orcidlink}

\begin{document}

\title{
One-loop Bern-Carrasco-Johansson  Numerators on Quadratic  Propagators
\\ from the Worldsheet   
}

\author{ Jin Dong \!\orcidlink{0000-0002-8747-7236}}%

\email{dongjin@itp.ac.cn}

\affiliation{CAS Key Laboratory of Theoretical Physics, Institute of Theoretical Physics, Chinese Academy of Sciences, Beijing 100190, China}
\affiliation{School of Physical Sciences, University of Chinese Academy of Sciences, No.19A Yuquan Road, Beijing 100049, China}

\author{ Yao{-}Qi Zhang \!\orcidlink{0000-0001-9211-1952}}%

\email{zhangyaoqi@itp.ac.cn}

\affiliation{CAS Key Laboratory of Theoretical Physics, Institute of Theoretical Physics, Chinese Academy of Sciences, Beijing 100190, China}
\affiliation{School of Physical Sciences, University of Chinese Academy of Sciences, No.19A Yuquan Road, Beijing 100049, China}

\author{Yong Zhang \!\orcidlink{0000-0002-3522-0885}}

\email{yzhang@perimeterinstitute.ca}

\affiliation{Perimeter Institute for Theoretical Physics, Waterloo, ON N2L 2Y5, Canada.}

\affiliation{
School of Physical Science and Technology, Ningbo University, Ningbo 315211, China.}

\begin{abstract}

We introduce a novel approach for deriving one-loop Bern-Carrasco-Johansson (BCJ) numerators and reveal the worldsheet origin of the 
one-loop 
double copy.
Our work shows that expanding Cachazo-He-Yuan  half-integrands into generalized Parke-Taylor factors intrinsically generates BCJ numerators on quadratic propagators satisfying Jacobi identities. We validate our methodology by successfully reproducing one-loop BCJ numerators for Non-Linear Sigma Model as well as those of pure Yang-Mills theory in four dimensions with all-plus or single-minus helicities.

\end{abstract}

\maketitle

\section{Introduction \label{sec1}}  

Recent advancements in quantum field theory have highlighted the essential role of scattering amplitudes in unraveling the fundamental interactions in nature. At the forefront of these advancements are the Bern-Carrasco-Johansson (BCJ) double-copy relations, rooted in the concept of color-kinematic duality ({cf.} \cite{Bern:2008qj,Bern:2010ue,Bern:2019prr,Bern:2022wqg,Adamo:2022dcm}). These relations have unveiled significant structures and simplifications in scattering amplitudes at both tree and loop levels, suggesting a complex interplay within Feynman diagrams that hints at a unified framework underlying gauge and gravitational theories.

Another noteworthy development in this area is the Cachazo-He-Yuan (CHY) formula \cite{Cachazo:2013gna,Cachazo:2013iea,Cachazo:2013hca,Cachazo:2014xea}, which offers an alternative approach to understanding scattering amplitudes beyond traditional Feynman diagram methods.   The CHY formula, known for its intricate geometry and combinatorial intelligence in the realm of worldsheet moduli space, streamlines the derivation of BCJ numerators at the tree level
and enhances the efficiency of amplitude calculations. Significant contributions in this field include the identification of relationships between various theories \cite{Cachazo:2014xea,Cheung:2017ems,Zhou:2018wvn} and the development of polynomial representations of BCJ numerators for Yang-Mills (YM) theory and many other theories of any multiplicity \cite{Du:2017kpo,Teng:2017tbo,Edison:2020ehu,He:2018pol}.

Moreover, ambitwistor strings \cite{Mason:2013sva,Geyer:2022cey}, along with traditional string theories \cite{Mafra:2011kj,Berkovits:2013xba,Adamo:2015hoa,Berkovits:2022ivl,Mafra:2022wml}, have provided deeper insights into the CHY formula from a world-sheet perspective and broadened the applicability of the CHY formula, particularly in the realm of loop amplitudes. Additionally, intersection theories have also proven instrumental to illuminate the mathematical and geometric foundations inherent in the CHY formula \cite{Mizera:2017rqa,Mizera:2019blq}.  It has also been established that manipulating tree-level CHY integrands through their forward limit in higher dimensions generates one-loop CHY integrands \cite{He:2015yua,Cachazo:2015aol}.

Despite considerable advancements, accessing BCJ numerators on quadratic propagators at loop level via worldsheet methods remains a daunting task, with the traditional worldsheet formula introducing loop propagators with linear loop momentum dependence, complicating analyses \cite{He:2015yua,Geyer:2015jch,Cachazo:2015aol,Geyer:2017ela,Edison:2020uzf}.  Although new BCJ double copies have been discovered employing these loop integrands \cite{He:2016mzd,He:2017spx}, the linear aspect hinders simplified integration.

However, significant research by Feng, He and two of the authors \cite{Feng:2022wee}, among many other works (cf. \cite{Baadsgaard:2015hia, Cardona:2016bpi, Cardona:2016wcr, Gomez:2016cqb, Gomez:2017lhy, Ahmadiniaz:2018nvr, Agerskov:2019ryp, Farrow:2020voh,Porkert:2022efy}), introduced a method for generating loop integrands with quadratic propagators, pivoting on the expansion of CHY half-integrands via one-loop generalized Parke-Taylor (PT) factors. This approach alleviates the complexities posed by linear dependencies in loop momentum, propelling forward the exploration of loop-level scattering amplitudes. 

In this paper, for the first time, we prove that the expansion onto generalized PT factors naturally gives rise to BCJ numerators on quadratic propagators that adhere to Jacobi identities using the formulas in \cite{Feng:2022wee}. 
Consequently, this significantly enhances  relevance of the CHY formulas, affirming their utility  at  the one-loop level. 
Besides, our method proposes a  more flexible  double-copy framework building upon the foundations established in previous works \cite{Bern:2008qj,Bern:2010ue}.

As a practical application of our methodology, we demonstrate its efficiency in reconstituting one-loop BCJ numerators for Non-Linear Sigma Model (NLSM) for arbitrary multiplicities  recently proposed in~\cite{Edison:2023ulf}   as well as those of pure YM theories in four dimensions with all-plus or single-minus helicity external gluons \cite{Boels:2013bi}. Simple double copies among them yield the loop integrands for special Galileon, Born-Infeld,  and pure gravity amplitudes \cite{Cachazo:2014xea} with corresponding helicities. 

\section{Quadratic
Propagators from the Worldsheet \label{sec2}}


\subsection{CHY Formula for One-loop Scattering Amplitudes}

\vspace{-.2cm}

The one-loop CHY formula yields loop integrands for the scattering of $n$ external, incoming massless particles by integrating over the moduli space of the degenerate tori, specifically, the nodal Riemann sphere localized by the one-loop scattering equations \cite{Geyer:2015bja,Geyer:2015jch}:
\ba
\label{oneloopchy}
\mathcal{M}(\ell){=}\frac{1}{\ell^2}
\int
\underbrace{\prod_{i=2}^{n} \mathrm{~d} \sigma_{i} \, \delta
\left(
\frac{2\,\ell\! \cdot\! k_{i} }{\sigma_{i}}{+}\sum_{\substack{j=1\\  j \neq i}}^{n} \frac{s_{ij} }{\sigma_{i j}}
\right)
}_{\equiv {\rm d}\mu_n}
\!
{\mathcal{I}}_{L}(\ell) {\mathcal{I}}_{R}(\ell)\,,
\non
\\[-6mm]
\ea
where $\sigma_{ij}\equiv\sigma_i-\sigma_j$ and $s_{ij}\equiv 2\, k_i\cdot k_j$. 
 These loop integrands can also be derived from $(n+2)$-point massless tree-level CHY formula via the forward limit in higher dimensions \cite{He:2015yua,Cachazo:2015aol}. Further integration upon the off-shell loop momentum $\ell$ yields the one-loop amplitudes. 
 
 We have fixed the ${\rm SL}(2,{\mathbb C})$ gauge redundancy in the nodal Riemann spheres in \eqref{oneloopchy} by setting the two nodal punctures as $\sigma_+\to 0, \sigma_-\to \infty$, and $\sigma_1\to 1$.  The measure ${\rm d} \mu _n$ is universal, while the half-integrands ${\mathcal{I}}_{L/R}(\ell)$ with equal ${\rm SL}(2,{\mathbb C})$ weights encode the dynamics information for a specific theory. The $n$ external momenta 
satisfy momentum conservation, $\sum_{i=1}^{n}k_i=0$.    For brevity, we introduce multiparticle momenta $k_{12 \ldots p}\equiv \sum_{i=1}^p k_i$ and the shorthand $\ell_{12\cdots p}\equiv \ell+k_{12\cdots p}$, such that  $\ell_{12\cdots p}^2$ signifies the quadratic loop propagator.

\subsection{One-Loop Cubic Graph and Quadratic Propagators}

An important ingredient in the one-loop double copy construction is the one-loop cubic graph \cite{Bern:2008qj,Bern:2010ue}. To describe this, we introduce a uniform notation, $ g({\cal A}_1, {\cal A}_2, \ldots, {\cal A}_m) $, where the sequence $ {\cal A}_1, {\cal A}_2, \ldots, {\cal A}_m $ with $ 1 \leq m \leq n  $ symbolizes all $m$  dangling trees located at the corners of the polygon and  the loop momentum $\ell$ is directed from ${\cal A}_m$ to ${\cal A}_1$ as illustrated below, 
\begin{align}
\non
\\[-.7cm] 
\label{defggraph}
\begin{aligned}
    g(\mathcal{A}_1,\mathcal{A}_2,\ldots,\mathcal{A}_m)
\end{aligned}\equiv
\begin{aligned}
	\begin{tikzpicture}[scale=1.3, decoration={
	markings,
	mark=at position 0.5 with { \arrow{stealth}}}]
	\node[regular polygon,minimum size = 2cm] (p) at (0,0) {};
	\draw[thick] (p.corner 1)--(p.corner 2); 
 \node at (0,-.45) {$\ell$};
 \draw[thick] (p.corner 2)--(p.corner 3);
	\draw[thick,postaction={decorate}] (p.corner 4)--(p.corner 3); 
	\draw[thick,dashed] (p.corner 4)--(p.corner 5);
	\draw[thick,dashed] (p.corner 5)--(p.corner 1);
	\draw[thick,fill=gray!50] (p.corner 4) circle ( 0.15 );
	\draw[thick,fill=gray!50] (p.corner 3) circle ( 0.15 );
	\draw[thick,fill=gray!50] (p.corner 2) circle ( 0.15 );
	\draw[thick,fill=gray!50] (p.corner 1) circle ( 0.15 );
	\draw[thick] (p.corner 4) node[right=10]{$\mathcal{A}_m$};
	\draw[thick] (p.corner 4) + (-50:0.15) -- ++(-50:0.3);
	\draw[thick] (p.corner 4) + (-130:0.15) -- ++(-130:0.3);
	\draw[thick] (p.corner 4)+(-90:0.15) node[below=0]{$\ldots$};
	\draw[thick] (p.corner 3) node[left=10]{$\mathcal{A}_1$};
	\draw[thick] (p.corner 3) + (-50:0.15) -- ++(-50:0.3);
	\draw[thick] (p.corner 3) + (-130:0.15) -- ++(-130:0.3);
	\draw[thick] (p.corner 3)+(-90:0.15) node[below=0]{$\ldots$};
	\draw[thick] (p.corner 2) node[left=15]{$\mathcal{A}_2$};
	\draw[thick] (p.corner 2) + (140:0.15) -- ++(140:0.3);
	\draw[thick] (p.corner 2) + (-140:0.15) -- ++(-140:0.3);
	\draw[thick] (p.corner 2)+(155:0.15) node[left]{$\vdots$};
	\draw[thick] (p.corner 1) + (50:0.15) -- ++(50:0.3);
	\draw[thick] (p.corner 1) + (130:0.15) -- ++(130:0.3);
	\draw[thick] (p.corner 1)+(90:0.15) node[above=0]{$\ldots$};
	\end{tikzpicture}
\end{aligned}
\,\,.
\end{align}
The exclusive use of cubic vertices allows for each dangling tree  ${\cal A}_i$ to be represented by a nested square bracket.

Additionally, we define $P_g$ as the product of all propagators in graph $g$, encompassing both loop and tree elements. For instance,
\begin{align}
\nonumber
P_{g([1,2],[[3,4],5],6)} \equiv P_{[1,2],[[3,4],5],6} = \ell^2 \ell^2_{12} \ell^2_{12345}  s_{12} s_{34} s_{345}\,.
\end{align}
Henceforth, for simplicity, when referencing a graph $g(\cdots)$ as a subscript, we drop both the $g$ symbol and the parentheses to declutter the notation.
Note that $P_g$ depends on the orientation and position of $\ell$:
\begin{align}
\label{cyclicp}
& P_{{\cal A}_m,{\cal A}_1,\cdots, {\cal A}_{m-1}}  = P_{{\cal A}_1,{\cal A}_2, \cdots, {\cal A}_m} \big|_{\ell\to \ell_{{\cal A}_m}}\,,
\\ &
\label{reverse}
P_{{\cal A}_1,{\cal A}_m, \cdots, {\cal A}_2} = P_{{\cal A}_1,{\cal A}_2, \cdots, {\cal A}_m} \big|_{\ell\to -\ell_{{\cal A}_1}}\,.
\end{align}

\subsection{Generalized Parke-Taylor Factors and Their Integrals}

Inspired by the maximally helicity violating  gluon amplitude formula \cite{Parke:1986gb}, tree-level PT factors are used in the tree-level CHY formula to encode the information of color ordering for theories like YM theory \cite{Cachazo:2013hca}. They only have simple poles and can act as the basis of the tree-level CHY half-integrands (cf. \cite{Aomoto87,Cachazo:2013iea,Cardona:2016gon,He:2018pol}). Building on this, the one-loop variant of PT factors was introduced in the one-loop CHY formula in \cite{Geyer:2015bja} to similarly convey color-ordering information.   In \cite{Feng:2022wee}, further operators acting on the standard scalar one-loop PT factors were introduced to define 
the generalized one-loop PT factors, 
\begin{align}\label{eq:pttensor}
&{\pmb \ell}_1^{\mu_1,\mu_2,\ldots,\mu_r}{\rm PT}(1,2, \cdots,n) 
{\equiv} 
\left( \prod_{j=1}^r {\pmb \ell}_1^{\mu_j}
\right)
{\rm PT}(1,2, \cdots,n)
\non
\\[-8mm]
\nonumber
\\
& {\equiv}  \sum_{i=1}^n 
\prod_{j=1}^r (\ell^{\mu_j} {-} k_{12\cdots i-1}^{\mu_j})
{\rm PT}^{\rm tree}({+},i,i{+}1,\cdots, i{-}1,{-}),
\non
\\[-5mm]
\end{align}
%
%
which have nontrivial dependency on the loop momentum \footnote{For example,
$
{\pmb \ell}_1^{\mu}{\rm PT}(1,2,3)$
$=
\ell^\mu {\rm PT}^{\rm tree}({+},1,2,3,{-})+
$
$(\ell^\mu\!{-}k_1^\mu)
{\rm PT}^{\rm tree}({+},\!2,3,1,{-})
{+}(\ell^\mu\!{-}k_{12}^\mu)
{\rm PT}^{\rm tree}({+},\!3
,1,2,{-})
$. }.
Leg 1 in \eqref{eq:pttensor} plays a special role as we will define $\ell$ as the loop momentum flowing into the subtree that contains leg 1 and we use the subscript in the operator ${\pmb \ell}_1$  to emphasize it. 
The tree-level PT factor reads
$
{\rm PT}^{\rm tree}({+},1,2,\cdots,n,{-})\equiv \frac{1}{\sigma_1 \sigma_{1,2} \sigma_{2,3} \cdots \sigma_{n-1, n}}
$.

It was demonstrated in \cite{Feng:2022wee}  that the CHY integral of two  generalized PT factors yields loop integrands with quadratic loop propagators,
\begin{align}
\label{twogeneralsym}
&\frac{1}{\ell^2}\int {\rm d}\mu_n\,
{\pmb \ell}_1^{\mu_1,\mu_2,\ldots,\mu_r}{\rm PT}\big(1,\rho(2), \cdots, \rho(n)\big)
\ln &
\qquad\qquad \times
{\pmb \ell}_1^{\nu_1,\nu_2,\ldots,\nu_t}{\rm PT}\big(1,\sigma(2), \cdots,\sigma(n)\big)
 \\
 \non
\cong& 
\,\,
{\rm sgn}^\rho_\sigma
\!\!\!
\sum_{
\substack
{
g\in T(1,\rho) \cap T(1,\sigma)
}}
\frac{
\ell^{\mu_1,\mu_2,\ldots,\mu_r}_{A(g,\rho)}
\ell^{\nu_1,\nu_2,\ldots,\nu_t} _{ A(g, \sigma)} 
}{ P_g}\,,
\non
\end{align}
where $\rho$ and $\sigma$ denote permutations of the elements $2,3,\cdots,n$ and the symbol $\cong$ signifies that the integrands on both sides yield identical results after loop integration ~\footnote{In \eqref{twogeneralsym}, we have omitted tadpoles since they are confined to scaleless integrals~\cite{Cachazo:2015aol,Bern:2013yya}, which do not contribute to the amplitudes. Massless bubbles like $g_{1,{\cal A}_2}$, which have formally been kept in \eqref{twogeneralsym}, can also be disregarded because of similar reasons. As a result, singular poles like $s_{12\cdots n}$ and $s_{2\cdots n}$, arising from the forward limit, do not pose any issues in the application of formulas \eqref{twogeneralsym}.}.
 The summation extends over all graphs that are members of both $T(1,\rho)$ and $T(1,\sigma)$  defined as follows.
 Considering a one-loop cubic graph  $g({\cal A}_1,{\cal A}_2,\cdots,{\cal A}_m)$ defined in \eqref{defggraph} with $2\leq m \leq n$, the loop momentum circulates clockwise and particle 1 is positioned in the initial corner, that is, $ {\cal A}_1 \ni 1 $, and the set $T(1,\rho)$ represents all of such cubic graphs with the external legs ordered according to the sequence $(1,\rho)$, in a clockwise arrangement. 

$\ell_{A(g,\rho)}^{\mu_1\mu_2}$ implies $\ell_{A(g,\rho)}^{\mu_1}\ell_{A(g,\rho)}^{\mu_2}$ and the shift factor $A(g,\rho)$ in $\ell_{A(g,\rho)}= \ell+k_{A(g,\rho)}$  signifies the subset of particles in the first corner ${\cal A}_1$ of graph $g$ that are situated before particle 1 in the cyclic order $(1,\rho)$. Note that $A(g,\rho)$ can be empty. More explicitly, suppose $\rho(b)\rho(b{+}1)\cdots \rho(n) 1 \rho(2)\cdots \rho(a)$ belong to ${\cal A}_1$; then, $A(g,\rho)=\{\rho(b),\cdots, \rho(n)\}$.
The overall sign  is delineated as follows,
\vspace{-.20cm}
\begin{align} 
\label{sign1}
{\rm sgn}^{\rho}_\sigma\equiv
\prod_{i=1}^{|\sigma|-1}{\rm sgn} ^{\rho}_{\sigma(i),\sigma(i+1)} ={\rm sgn}_{\rho}^\sigma \,,
\end{align} 
where ${\rm sgn}^{\rho}_{i,j}$ equals $+1$ if $i$ is ahead of $j$ in $\rho$, and $-1$ otherwise.

In the next section, we will show how to derive BCJ numerators on quadratic propagators based on \eqref{twogeneralsym}.

\section{One-loop BCJ Numerators from the Worldsheet}\label{sec3}

In the study of one-loop CHY formulas for theories such as those with $SU(N)$ or $SO(N)$ color groups, the formulation involves distinctive half-integrands. The first half-integrand, ${\mathfrak C}_n$, represents a color-dressed scalar PT  factor, devoid of all kinematic considerations, defined as
\ba
\label{defineCn}
& {\mathfrak C}_n\equiv  \sum_{\sigma \in S_{n-1}} { C}_{1,\sigma} {\rm PT}(1,\sigma),
\ln & \qquad {\rm where} ~ { C}_{1,2, \ldots ,n} \equiv f^{z a_{1} b} f^{b a_{2} c} f^{c a_{3} d} \ldots f^{y a_{n} z}\,,
\ea
with $f^{abc}$ representing the structure constants of the color group.

The second half-integrand, $I_n$, is in general more complicated. However, as we prove later, for theories that accept a BCJ double copy, their second half-integrand can always be expanded to generalized PT factors \eqref{eq:pttensor}. 
This expansion can be organized based on scalar PT factors ${\rm PT}(1,\rho)$, with all operators ${\pmb \ell}_1$ and $\sigma$-independent variables like $k_i$, polarizations $\epsilon_i$, { etc.}, consolidated as a single operator ${\pmb N}_{1,\rho}$. Consequently, $I_n$ mirrors the structure of ${\mathfrak C}_n$, described as
%
\begin{align}
\label{defIn}
I_n= \sum_{\rho \in S_{n-1}} {\pmb N}_{1,\rho} {\rm PT}(1,\rho)\,.
\end{align}

In the next section, we  show the concrete expansions \eqref{defIn} for NLSM and pure YM theory with all-plus or single-minus helicities. More examples of the expansions for low-point 
super-Yang-Mills amplitudes can be found in  \cite{Feng:2022wee},  suggesting a potential for generalization to higher-point scenarios \cite{Edison:2021ebi,Mafra:2018pll,Mafra:2018qqe}. In this section, our focus is on using the (abstract) expansion \eqref{defIn} as a foundation to illustrate a universal approach for deriving BCJ numerators.

\subsection{General Claim\label{secdoublecyopy}}

Our central statement is that the integral $\frac{1}{\ell^2}\int {\rm d}\mu_n I_n {\mathfrak C}_n$ inherently generates BCJ numerators for theory ${\cal O}$. Specifically, the master BCJ numerator for an $n$-gon graph can be straightforwardly acquired via the substitution
\begin{align}
\label{masterbcjnu}
{ N}_{1, \rho(2),\cdots, \rho(n)} \equiv { N}_{g(1, \rho(2),\cdots, \rho(n))} = {\pmb N}_{1,\rho} \big|_{{\pmb \ell}_1\to \ell }\,.
\end{align}

{\bf{Jacobi identities }} For any triplet graphs with identical placement and orientation of loop momentum $\ell$ but differing by a single propagator (as shown below), 
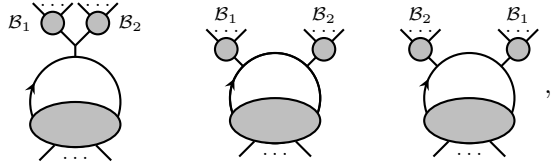
\begin{equation*}
\begin{aligned}
\begin{tikzpicture}[decoration={
markings,
mark=at position .57 with { \arrow{stealth}}},scale=1.5]
\draw[thick,postaction={decorate}] (3.9,0) arc (360:0:0.4);
\draw[thick,fill=gray!50] (0+3.5,-0.2) ellipse (0.4 and 0.2 );
\draw[thick] (3.5,0.4)-- ++(0,0.1);
\draw[thick] (3.5,0.5)-- ++(-0.2,0.2);
\draw[thick] (3.5,0.5)-- ++(0.2,0.2);
\draw[thick,fill=gray!50] (3.3,0.7) circle ( 0.1 );
\draw[thick,fill=gray!50] (3.7,0.7) circle ( 0.1 );
\draw[thick] (3.3,0.7)+(135:0.1)-- ++(135:0.25) ;
\draw[thick] (3.3,0.7)+(45:0.1)-- ++(45:0.25) ;
\draw[thick] (3.3,0.7+0.16) node{\scriptsize $\ldots$};
\draw (3.25-0.05,0.7) node[left=0pt]{\scriptsize {$\mathcal{B}_1$}};
\draw[thick] (3.7,0.7)+(135:0.1)-- ++(135:0.25) ;
\draw[thick] (3.7,0.7)+(45:0.1)-- ++(45:0.25) ;
\draw[thick] (3.7,0.7+0.16) node{\scriptsize$\ldots$};
\draw (3.75+0.05,0.7) node[right=0pt]{\scriptsize {$\mathcal{B}_2$}};
\draw[thick] (+3.5,-.2)+(225:0.26)-- ++(225:0.45); 
\draw[thick] (+3.5,-.2)+(315:0.26)-- ++(315:0.45);
\draw[thick] (+3.5,-.2)+(270:0.2) node[below=0pt]{\scriptsize$\ldots$};
\end{tikzpicture}
\end{aligned}
\qquad
\begin{aligned}
\begin{tikzpicture}[decoration={
			markings,
			mark=at position 0.57 with { \arrow{stealth}}},scale=1.5]
\draw[thick,postaction={decorate}] (0.4,0) arc (360:0:0.4);
\draw[thick] (-.4+.4, 0.0) circle ( 0.4 );
		\draw[thick,fill=gray!50] (0,-.2) ellipse (0.4 and 0.2 );
\draw[thick] (-0.282843,0.282843)-- ++(-0.15,0.15);
\draw[thick,fill=gray!50] (-0.282843-0.15,0.282843+0.15) circle ( 0.1 );
\draw[thick] (0.282843,0.282843)-- (0.282843+0.15,0.282843+0.15);
\draw[thick,fill=gray!50] (0.282843+0.15,0.282843+0.15) circle ( 0.1 );
\draw[thick] (-0.282843-0.15,0.282843+0.15)+(135:0.1)-- ++(135:0.25) ;
\draw[thick] (-0.282843-0.15,0.282843+0.15)+(45:0.1)-- ++(45:0.25) ;
\draw[thick] (-0.282843-0.15,0.282843+0.25)+(0,0.06) node {\scriptsize$\ldots$};
\draw[thick] (-0.282843-0.15,0.282843+0.25)+(0,0.2) node {\scriptsize $\mathcal{B}_1$};
\draw[thick] (0.282843+0.15,0.282843+0.15)+(135:0.1)-- ++(135:0.25) ;
\draw[thick] (0.282843+0.15,0.282843+0.15)+(45:0.1)-- ++(45:0.25) ;
\draw[thick] (0.282843+0.15,0.282843+0.25)+(0,0.06) node {\scriptsize$\ldots$};
\draw[thick] (0.282843+0.15,0.282843+0.25)+(0,0.2) node {\scriptsize $\mathcal{B}_2$};
\draw[thick] ($(225:0.26)+(0,-.2)$)-- ($(225:0.45)+(0,-.2)$); 
\draw[thick] ($(315:0.26)+(0,-.2)$)-- ($(315:0.45)+(0,-.2)$);
\draw[thick] (270:0.2)+(0,-.2) node[below=0pt]{\scriptsize$\ldots$};
\end{tikzpicture}
\end{aligned} 
\qquad
\begin{aligned}
	\begin{tikzpicture}[decoration={
			markings,
			mark=at position 0.57 with { \arrow{stealth}}},scale=1.5]
%
\draw[thick,postaction={decorate}] (0.4,0) arc (360:0:0.4);
		\draw[thick,fill=gray!50] (0,-.2) ellipse (0.4 and 0.2 );
\draw[thick] (-0.282843,0.282843)-- ++(-0.15,0.15);
\draw[thick,fill=gray!50] (-0.282843-0.15,0.282843+0.15) circle ( 0.1 );
\draw[thick] (0.282843,0.282843)-- (0.282843+0.15,0.282843+0.15);
\draw[thick,fill=gray!50] (0.282843+0.15,0.282843+0.15) circle ( 0.1 );
\draw[thick] (-0.282843-0.15,0.282843+0.15)+(135:0.1)-- ++(135:0.25) ;
\draw[thick] (-0.282843-0.15,0.282843+0.15)+(45:0.1)-- ++(45:0.25) ;
\draw[thick] (-0.282843-0.15,0.282843+0.25)+(0,0.06) node {\scriptsize$\ldots$};
\draw[thick] (-0.282843-0.15,0.282843+0.25)+(0,0.2) node {\scriptsize $\mathcal{B}_2$};
\draw[thick] (0.282843+0.15,0.282843+0.15)+(135:0.1)-- ++(135:0.25) ;
\draw[thick] (0.282843+0.15,0.282843+0.15)+(45:0.1)-- ++(45:0.25) ;
\draw[thick] (0.282843+0.15,0.282843+0.25)+(0,0.06) node {\scriptsize$\ldots$};
\draw[thick] (0.282843+0.15,0.282843+0.25)+(0,0.2) node {\scriptsize $\mathcal{B}_1$};
\draw[thick] ($(225:0.26)+(0,-.2)$)-- ($(225:0.45)+(0,-.2)$); 
\draw[thick] ($(315:0.26)+(0,-.2)$)-- ($(315:0.45)+(0,-.2)$);
\draw[thick] ($(270:0.2)+(0,-.2)$) node[below=0pt]{\scriptsize$\ldots$};
\end{tikzpicture}
\end{aligned} \,,
\end{equation*}
their numerators satisfy Jacobi identities similar to color factors. This denotes an antisymmetrization of the numerators,
\begin{align} 
\label{numjac}
N_{\cdots, [{\cal B}_1,{\cal B}_2], \cdots} = N_{\cdots, {\cal B}_1,{\cal B}_2, \cdots} - N_{\cdots, {\cal B}_2,{\cal B}_1, \cdots},
\end{align}
where the ellipsis represents consistent dangling tree sequences across the three graphs.

Applying these identities recursively enables deriving the numerator ${ N}_g$ of any graph as a linear combination of $(n-1)!$ master BCJ numerators ${ N}_{1,\rho}$ with $\rho \in S_{n-1}$, considering potential loop momentum shifts.

The shift arises because varying $\ell$ positions within the same $n$-gon yield different numerator representations,
\ba
\label{shiftell}
{ N}\!\left(
\!\!\!\!
\begin{aligned}
\resizebox{2.8cm}{1.4cm}{
\begin{tikzpicture}[scale=0.5,decoration={
		markings,
		mark=at position 0.5 with {\arrow{stealth}}}]
\draw[thick] (0,0) circle (0.7 );
\draw [postaction={decorate}] (-90:0.7) arc (270:90:0.7);
\draw (180:0.7) node[left=0pt]{\scriptsize $\ell$};
\draw (90:0.7) node[above=0pt]{\scriptsize $\ldots$};
\draw (270:0.7) node[below=0pt]{\scriptsize $\ldots$};
\draw[thick] (45:0.7)--++(45:0.6) ;
\draw (45:1.3) node[right=-2pt]{\scriptsize $\gamma(i)$};
\draw[thick] (-45:0.7)--++(-45:0.6) ;
\draw (-45:1.3) node[right=-2pt]{\scriptsize$\gamma(i{+}1)$};
\draw[thick] (135:0.7)--++(135:0.6) ;
\draw (135:1.3) node[left=-2pt]{\scriptsize $\gamma(2)$};
\draw[thick] (-135:0.7)--++(-135:0.6) ;
\draw (-135:1.3) node[left=-2pt]{\scriptsize $\gamma(n)$};
\draw[thick] (0:0.7)--++(0:0.4) ;
\draw (0:1.1) node[right=-2pt]{\scriptsize $1$};
\end{tikzpicture}}
\end{aligned} 
\!\!
\right)\! {\equiv} \,
{ N}
\!\left(
\!\!\!\!
\begin{aligned}
\resizebox{2.8cm}{1.4cm}{
\begin{tikzpicture}[scale=0.5,decoration={
		markings,
		mark=at position 0.5 with {\arrow{stealth}}}]
\draw[thick] (0,0) circle (0.7 );
\draw [postaction={decorate}] (90:0.7) arc (90:-60:0.7);
\draw (45:0.3) node{\scriptsize $\ell$};
\draw (90:0.7) node[above=0pt]{\scriptsize $\ldots$};
\draw (270:0.7) node[below=0pt]{\scriptsize $\ldots$};
\draw[thick] (45:0.7)--++(45:0.6) ;
\draw (45:1.3) node[right=-2pt]{\scriptsize $\gamma(i)$};
\draw[thick] (-45:0.7)--++(-45:0.6) ;
\draw (-45:1.3) node[right=-2pt]{\scriptsize $\gamma(i{+}1)$};
\draw[thick] (135:0.7)--++(135:0.6) ;
\draw (135:1.3) node[left=-2pt]{\scriptsize $\gamma(2)$};
\draw[thick] (-135:0.7)--++(-135:0.6) ;
\draw (-135:1.3) node[left=-2pt]{\scriptsize $\gamma(n)$};
\draw[thick] (0:0.7)--++(0:0.4) ;
\draw (0:1.1) node[right=-2pt]{\scriptsize $1$};
\end{tikzpicture}
}
\end{aligned} 
\!\!\!\right) _{
\ell\to \ell_{\gamma(2),\ldots,\gamma(i) }} 
\!\!\!\!\!\!\!\!\!\!\!\!\!\!.
\qquad\qquad\qquad\qquad\qquad
\non
\\[-1.18cm]
\\[0.05cm]
\non
\ea 
That is ${ N}_{\gamma(2),\ldots, \gamma(i), 1 , \ldots } \equiv { N}_{1, \ldots, \gamma(2),\ldots, \gamma(i) }\big|_{\ell\to \ell_{ \gamma(2),\ldots ,\gamma(i) }}$, where $\gamma(2)$,$\gamma(3), \ldots, \gamma(n)$ denote a permutation of $2,3,\ldots,n$.

\vspace{.15cm}

{\bf{Double copy }} 
Having obtained all BCJ numerators and color factors for the one-loop cubic graphs, we  express
the loop integrand for theory ${\cal O}$ as
\begin{equation}
\label{proposal}
\frac{1}{\ell^2}\int {\rm d}\mu_n \, I_n \,{\mathfrak C}_n \cong {\cal M}^{{\cal O}}_n(\ell) = \sum_{g \in W_1} \frac{{ N}_g C_g}{P_g},
\end{equation}
where $W_1$ denotes all one-loop cubic graphs $g({\cal A}_1,{\cal A}_2,\cdots,{\cal A}_m)$ with $2\leq m \leq n$ and 
$ {\cal A}_1   \ni 1 $.

Consider another theory ${\cal O}'$ with half-integrand $I'_n= \sum_{\sigma \in S_{n-1}} {\pmb N}'_{1,\sigma} {\rm PT}(1,\sigma)$. The  double copy between ${ \cal O}$ and ${\cal O}'$ gives the loop integrand of a third distinct theory,
\vspace{-.15cm}
\begin{align}
\label{proposal2}
\frac{1}{\ell^2}\int {\rm d}\mu_n \, I_n \,I'_n \cong {\cal M}^{\rm DC}_n(\ell) = \sum_{g\in W_1} \frac{N_g N^\prime_g}{P_g},
\end{align}
%
where $C_g$ in \eqref{proposal} is replaced by another set of BCJ numerators $N^\prime_g= N_g|_{{\pmb N}\to {\pmb N}'}$.

\vspace{.15cm}

{\bf{ Numerators }} 
 For clarity, the numerator $N_g$ in \eqref{proposal} and \eqref{proposal2} is explicitly formulated as
%
\begin{align} 
\label{ngexpre}
{ N}_g = {\rm sgn}^{\rho_0}_g \sum_{ (1,\rho)\in T^{-1}(g)} {\rm sgn}^{\rho_0}_{\rho}  \, {\pmb N}_{1,\rho}\big|_{ {\pmb \ell}_1\to \ell_ {A(g,\rho)}}\,.
\end{align}
Here, the sum is over all orderings $(1,\rho)$ such that  $g\in T(1,\rho)$ and we have selected an arbitrary $(1,\rho_0)$ from them as a reference.  The 
sign ${\rm sgn}^{\rho_0}_{\rho}$ is specified in \eqref{sign1}.

In a parallel manner, the color factor is given by
\vspace{-.15cm}\!\!\!
\begin{align}
\label{cgexpre2}
{C}_g ={\rm sgn}^{\rho_0}_g \sum_{ (1,\sigma)\in T^{-1}(g)} {\rm sgn}^{\rho_0}_{\sigma}  \, C_{1,\sigma}\,.
\end{align}
Notably, the comprehensive overall signs ${\rm sgn}^{\rho_0}_g$ in \eqref{ngexpre} and \eqref{cgexpre2} result in $+1$ upon the multiplication of $N_g$ and $C_g$ in \eqref{proposal}, ensuring consistency.

{\bf Example} Here is an example at $n=3$ for \eqref{proposal}:
\ba
\label{combiIn3ptc}
\frac{1}{\ell^2}\int & {\rm d}\mu_3 \, I_3 \,{\mathfrak C}_3
\cong
\frac{N_{1,2,3} C_{1,2,3}}{P_{1,2,3}}+  \frac{N_{1,3,2} C_{1,3,2}}{P_{1,3,2}}
 \\ 
 \non & + \frac{N_{1,[2,3]} C_{1,[2,3]}}{P_{1,[2,3]}}+ \frac{N_{[1,2],3} C_{[1,2],3}}{P_{[1,2],3}}+\frac{N_{[1,3],2} C_{[1,3],2}}{P_{[1,3],2}}\,,
\ea 
where $ 
N_{[1,2],3}\!=\!N_{1,2,3}\!-\!N_{2,1,3}\!= \!{\pmb N}_{1,2,3} \big|_{{\pmb \ell}_1\to \ell } \!-\!{\pmb N}_{1,3,2} \big|_{{\pmb \ell}_1\to \ell_2 }  $. 
The example of \eqref{proposal2} can be easily given by substituting $C$ with $N'$.

\subsection{Proof}

We proceed to prove our general proposal as outlined in \eqref{proposal2}. Subsequently, \eqref{proposal} is affirmed as a corollary, wherein $C_{1,\sigma}$ can be treated as a specific case of ${\pmb N}'_{1,\sigma}$ that does not involve ${\pmb \ell}_1$.

First, we examine the integral of ${\pmb N}_{1,\rho} {\rm PT}(1,\rho)$ with ${\pmb N}'_{1,\sigma}{\rm PT}(1,\sigma)$. In general, ${\pmb N}_{1,\rho}$ is a linear combination of ${\pmb \ell}_1^{\mu_1,\mu_2,\ldots,\mu_{r}}$ with varying rank $r$
\vspace{-.15cm}\!\!\!
\ba 
{\pmb N}_{1,\rho}=  \sum_{r=0}^\infty M_{\rho, r} {\pmb \ell}_1^{\mu_1, \mu_2,\cdots, \mu_{r}}\,,
\ea 
where $M_{\rho, r}$ are coefficients free of ${\pmb \ell}_1$.
 According to \eqref{twogeneralsym}, the shift factor $k_{A(g,\rho)}$ is universal across all ranks $r$ of operator ${\pmb \ell}_1^{\mu_1,\mu_2,\ldots,\mu_{r}}$. This results in a universal shift for the entire operator ${\pmb N}_{1,\rho}$ for any $g$ and $\rho$,
 \vspace{-.15cm}\!\!\!
 \ba 
 \sum_{r=0}^\infty \!\Big(\! M_{\rho, r} {\pmb \ell}_1^{\mu_1, \mu_2,\cdots, \mu_{r}} \!\big|_{ {\pmb \ell}_1\to \ell_ {A(g,\rho)}}\!\Big) \!=\!  {\pmb N}_{1,\rho} \big|_{ {\pmb \ell}_1\to \ell_ {A(g,\rho)}}.
 \ea 
 The same applies to ${\pmb N}'_{1,\sigma}$, simplifying the integral to
 \vspace{-.15cm}\!\!\!
\begin{align}
\label{twogeneralsymcolor2}
&\frac{1}{\ell^2}\int {\rm d}\mu_n
{\pmb N}_{1,\rho} {\rm PT}(1,\rho) 
{\pmb N}'_{1,\sigma} {\rm PT}(1,\sigma)
 \\
 \non
\cong& 
\,\,
{\rm sgn}^\rho_\sigma
\!\!\!\!\!\!
\!\!\!
\sum_{
\substack
{
g\in T(1,\rho) \cap T(1,\sigma)
}}
\!\!\!
\!\!\!
\!\!\!
\frac{
{\pmb N}_{1,\rho}\big|_{ {\pmb \ell}_1\to \ell_ {A(g,\rho)}}
{\pmb N}'_{1,\sigma}\big|_{ {\pmb \ell}_1\to \ell_ {A(g,\sigma)}}
}{ P_g}\,. 
\non
\end{align}
By summing over all $\rho$ in $I_n$ defined in \eqref{defIn} and all $\sigma$ in $I'_n$, and consolidating the contributions for each graph, we can deduce \eqref{proposal2} and, subsequently, \eqref{proposal}. Additional information on signs, Jacobi identities \eqref{numjac}, tadpole contributions, and  more examples are provided in \cref{appendixa}.

\subsection{Refined Double Copy \label{sec4}}

In our construction \eqref{proposal} and \eqref{proposal2}, graph pairs $g({\cal A}_1,{\cal A}_2, \ldots, {\cal A}_m)$ and $g({\cal A}_1,{\cal A}_m, \ldots, {\cal A}_2)$, typically seen as identical for $3 \leq m \leq n$, are distinct in set $W_1$. We introduce set ${\hat W}_1$ by merging such pairs in $W_1$ and formulate a new CHY half-integrand tied to $I_n$ as follows,
\begin{align}
\label{defInbar}
& {\bar I}_n = \sum_{\rho \in S_{n-1}} {\bar {\pmb N}}_{1,\rho} {\rm PT}(1,\rho), 
\\ 
\nonumber
& \qquad {\rm with} ~ {\bar {\pmb N}}_{1,\rho} = \frac{1}{2}
\big({ {\pmb N}}_{1,\rho} + (-1)^n { {\pmb N}}_{1,\rho^T} \big|_{{\pmb \ell}_1\to -{\pmb \ell}_1 - k_1} \big),
\end{align}
where $\rho^T$ reverses $\rho$. Employing ${\bar I}_n$ in place of $I_n$ in the CHY integral \eqref{proposal} reveals new master BCJ numerators ${\bar N}_{1,\rho}=\left.{\bar{\pmb N}}_{1,\rho}\right|_{{\pmb \ell}_1\to\ell}$ satisfying 
\ba 
\label{requirments2}
{\bar N}_{1,\rho^T}= (-1)^n\, {\bar N}_{1,\rho}\big|_{\ell \to -\ell-k_1}.
\ea 
This ensures identical contributions from $n$-gon pairs $g(1,\rho)$ and $g(1,\rho^T)$ to amplitudes, extending to any $g({\cal A}_1,{\cal A}_2, ..., {\cal A}_m)$ and $g({\cal A}_1,{\cal A}_m, ..., {\cal A}_2)$ as proved in \cref{appendixb}. Importantly, the new loop integrand matches $M(\ell)$ from \eqref{proposal} after integration, establishing ${\bar I}_n \cong I_n$ as a refined $I_n$, leading to the refined double copies,
\ba 
\label{proposal22}
{\cal M}^{{\cal O}}_n(\ell)\! \cong  \!\! \sum_{g \in {\hat W}_1} 
\!\! {\cal S}_g\frac{{\bar N}_g C_g}{P_g}  \,, \, {\cal M}^{\rm DC}_n(\ell) \! \cong \!\! \sum_{g\in {\hat W}_1 } \!\! {\cal S}_g  \frac{{\bar  N}_g {\bar N}'_g}{P_g} \,. 
\ea 
Here,  the symmetry factor ${\cal S}_g$ is 1 for bubble graphs but becomes 2 for triangles and larger polygons \footnote{ Here our symmetry factor relates to the conventional one in \cite{Bern:2008qj,Bern:2010ue} by ${\cal S}_g = 2/ S_g $ }. 

In particular, our initial double copy \eqref{proposal} and \eqref{proposal2}, which naturally arises from worldsheet perspectives, does not mandate \eqref{requirments2}, allowing unrelated BCJ numerators for identical graphs with differing $\ell$ orientations. 
This suggests potential redundancies in loop-level double-copy constructions, urging further exploration into higher-loop BCJ numerators.

In the context of scattering of $n$ identical external bosons, we achieve one-loop crossing-symmetric  BCJ numerators \cite{Broedel:2011pd,Carrasco:2011mn,Bern:2015ooa,Edison:2022jln} through averaging permutations of particle labels in the half-integrand ${\bar I}_n$, as delineated in \cref{sec5}.

 Having established the derivation of master BCJ numerators from the CHY half-integrand expansion, we posit that the inverse is equally valid.
  When provided with BCJ numerators that satisfy the Jacobi identities \eqref{numjac}, 
we can elevate the master numerators symbolized as $N(1,\rho)$ for $n$-gons  to operators,
$
N(1,\rho) \big|_{\ell \to {\pmb {\ell }}_1} \to {\pmb N}_{1,\rho}\,.
$
The CHY half-integrand for theory ${\cal O}$ then follows the expansion \eqref{defIn}, inherently yielding the same loop integrand for theory ${\cal O}$ according to \eqref{proposal}. As a corollary, this proves the existence of the expansion \eqref{defIn} for theories that accept BCJ double copies.

Further exploration in this area reveals additional insights.  
In \eqref{oneloopchy}, we have assumed that the CHY integrand for a given theory decomposes into two half-integrands, $\mathcal{I}_{L}(\ell)$ and $\mathcal{I}_{R}(\ell)$; however, it could in principle just be a quadratic combination of them, $\sum_{i} \mathcal{I}_{L}^{(i)}(\ell)\mathcal{I}^{(i)}_{R}(\ell)$. Nevertheless, our investigations confirm that the existence of BCJ numerators for a theory ensures that its CHY integrands can indeed be represented as a product of two half-integrands.

\section{Applications\label{sec:app}}

In this section, we demonstrate our approach through an analysis of one-loop amplitudes for pure gluons in $D=4$, focusing on all-positive or single-minus helicity configurations. Using straightforward techniques, similar to those employed at the tree level \cite{Cachazo:2014xea,Cheung:2017ems,Zhou:2018wvn}, we successfully derive the BCJ numerators for NLSM theories as well. 
This not only demonstrates the straightforwardness of the method in deriving BCJ numerators from the worldsheet but also indicates its broad applicability to diverse particle types.

Utilizing spinors, we can express the four-dimensional polarizations $\epsilon_i$ and momenta $k_i$, ensuring $\epsilon_i\cdot\epsilon_j \to 0$ via a specific reference spinor choice \cite{elvang2014scattering,schubert2014scattering}.
For all-plus or single-minus helicity configurations in gluon scattering, as established in \cite{Grisaru:1979re,Bern:1993wt},  loop amplitudes involving  a gluon, fermion, or scalar in the loop are proportionally equivalent. Thus, we concentrate exclusively on the scalar case.

We derive the corresponding one-loop CHY half-integrand from the tree level in the forward limit \cite{He:2015yua,Edison:2020uzf}. The tree-level ones for $n$ external gluons and two scalars, denoted as $+,-$, can be straightforwardly obtained from the well-known reduced Pfaffian for $n+2$ external gluons \cite{Cachazo:2013iea,Cachazo:2013hca} by extracting its coefficient $\epsilon_{+}\cdot \epsilon_{-}$ \cite{Cachazo:2014xea,Cheung:2017ems}. When setting all remaining $\epsilon_i\cdot\epsilon_j \to 0$, the Pfaffian simplifies to a determinant.
Implementing the forward limit yields the one-loop CHY half-integrand for $n$ external gluons with a scalar propagating in the loop,
\vspace{-.15cm}\!\!\!
\begin{equation}\label{eq:Psiexp2}
    I_n^{\rm YM}=- {\rm det} \, {\texttt C}(\epsilon_i,k_i,\sigma_i),
\end{equation}
where ${\texttt C}$ is an $n\times n$ matrix
defined as
\vspace{-.15cm}\!\!\!
\ba 
\label{cmatrix}
{\texttt C}_{i,j}\equiv \frac{\epsilon_i\cdot k_j}{\sigma_{i,j}} \, {\rm  for} ~i\neq j 
\, {\rm and}~  {\texttt C}_{i,i}\equiv \! -\!\!\!\!\! \sum_{\substack{a=1
,
a\neq i}}^n \!\!\!\!
{\texttt C}_{i,a} \!- \!\frac{\epsilon_i\cdot \ell}{\sigma_{i}}\,.
\ea 
Applying the matrix tree theorem \cite{Feng:2012sy} to expand ${\rm det} \, {\texttt C}$ results in a summation over all labeled trees $G$ with nodes $\{+,1,2,\cdots ,n\}$ and orientations of the $n$ edges $e(i,j)$ flowing to the root node $+$ \cite{Gao:2017dek} \footnote{For examples, for $n=4$, there are labelled trees like $(4\to 2; 3\to2\to 1\to +)$ whose contribution is given by $-\frac{\epsilon_4\cdot k_2}{\sigma_{4,2}}\frac{\epsilon_3\cdot k_2}{\sigma_{3,2}}\frac{\epsilon_2\cdot k_1}{\sigma_{2,1}}\frac{\epsilon_1\cdot \ell}{\sigma_{1}}$.  },
\vspace{-.15cm}\!\!\!
\ba 
\label{eq26eq26}
    I_n^{\rm YM}  = (-1)^{n+1} \sum_{G} \prod_{e(i,j)}  \frac{\epsilon_i\cdot k_j}{\sigma_{i,j}} \,.
\ea 
By utilizing partial fraction identities and grouping coefficients for each $\mathrm{PT^{tree}}({+},\pi,{-})$ \cite{He:2021lro}, we express \eqref{eq26eq26} as
\vspace{-.15cm}\!\!\!
\begin{equation}\label{eq:Psiexp}
    I_n^{\rm YM}=\sum_{\pi\in S_{n}}\prod_{i=1}^{n}\epsilon_i\cdot(\ell+Y_i(\pi)) \mathrm{PT^{tree}}(+,\pi ,-)\,,
\end{equation}
where $Y_i(\pi(1),\cdots,$ $\pi(j),i$ $,\cdots)$ $= k_{\pi(1),\cdots,\pi(j)}$.

As proved in \cref{appendixd}, $I_n^{\rm YM}$ can be expanded to the generalized PT factors 
 with ranks ranging from $2$ to $n$,
 \vspace{-.15cm}\!\!\!
\begin{equation}\label{eq:psi1lf}
  I_n^{\rm YM}=\sum_{\rho\in S_{n-1}}\prod_{i=1}^n\epsilon_i\cdot ({\pmb \ell}_1+Y_i(1,\rho))\mathrm{PT}(1,\rho)\,.
\end{equation}
Utilizing \eqref{masterbcjnu}, we derive all master BCJ numerators as
\vspace{-.15cm}\!\!\!
\begin{align} 
\label{ymnumerator}
N^{\rm YM}_{1,\rho}= \prod_{i=1}^n\epsilon_i\cdot (\ell+Y_i(1,\rho))\,.
\end{align}
One can easily check that $N^{\rm YM}_{1,\rho^T}=(-1)^n N^{\rm YM}_{1,\rho}\big|_{\ell\to -\ell_1}$. Consequently, we directly employ \eqref{proposal22} and \eqref{ngexpre} to compute the loop integrands for all-plus or single-minus YM and GR amplitudes.

Note that we derive \eqref{ymnumerator} directly from ${\rm det}{\texttt C}(\epsilon_i,k_i,\sigma_i)$ \footnote{ One can also derive \eqref{eq:Psiexp} from the polynomial representation of BCJ numerators for Yang-Mills-scalar theories as proposed in \cite{Fu:2017uzt,Du:2017gnh,Schlotterer:2016cxa,He:2019drm} upon setting all $\epsilon_i\cdot\epsilon_j \to 0$, but our methodology is more straightforward.}, ensuring manifest symmetry across all $n$ external gluons. Consequently, the master BCJ numerators, represented as ${{N}}^{\rm YM}_{1,2,\cdots,n} = \prod_{i=1}^n\epsilon_i\cdot \ell_{12\cdots i-1}$, exhibit crossing symmetry, as elucidated in \cref{sec5}, in scenarios involving all-plus helicities.

Remarkably, it is straightforward to extend our derivation from gluons to pions. Starting at the one-loop CHY half-integrand \eqref{eq:Psiexp2} for YM theory with a scalar running in the loop, simply replacing $\epsilon_i \to k_i$ in \eqref{eq:Psiexp2}, we get the half-integrand for NLSM~\cite{Cachazo:2014xea} in the one-loop CHY formula. Crucially,  we have not taken the the derivative of Lorentz product of a pair of external polarizations in this procedure which is different 
from the one at tree level~\cite{Cachazo:2014xea,Cheung:2017ems,Cheung:2017yef}.
Then a parallel derivation further demonstrates that  ${{N}}^{\rm NLSM}_{1,2,\cdots,n} = \prod_{i=1}^n k_i\cdot \ell_{12\cdots i-1}$. The double copies of themselves or together with the gluon ones \eqref{ymnumerator} produce the loop integrands for special Galileon and Born–Infeld theories with corresponding helicities, which will be further studied in \cite{nlsmtoappear}.

Although the BCJ numerators \eqref{ymnumerator} for gluons have been previously
presented in \cite{Pavao:2022kog,Geyer:2017ela,Boels:2013bi,Bern:1993wt,Bern:1998sv,Bern:1996ja} and those of pions recently  proposed in~\cite{Edison:2023ulf}, our approach reproduces
them universally in a streamlined and elegant
manner. 
The success in these specific cases highlights the adaptability of our method, suggesting its potential applicability to a broader array of theories.

\section{Discussion}

This research signifies an important advance in quantum field theories, particularly in computing one-loop BCJ numerators. We have pioneered a method for extracting one-loop BCJ numerators,  generating one-loop integrands using CHY half-integrands expanded into generalized PT factors.
Our strategy, tested on Non-Linear Sigma Model  as well as pure Yang-Mills theories with all-plus or single-minus helicities, demonstrates robustness and flexibility, showing great potential to enhance computational techniques and uncover previously unknown connections among one-loop amplitudes. See \cite{Xie:2024pro} for recent progress. 

By expressing any one-loop double copy as a CHY integral combining a direct product of two half-integrands, our method potentially alludes to a one-loop variant of the Kawai-Lewellen-Tye (KLT) relations \cite{Kawai:1985xq,Cachazo:2013gna,Cachazo:2016sdc} using quadratic propagators (see \cite{He:2016mzd,He:2017spx}  using linear propagators), potentially linked to recent studies on the genus-one KLT relations in string amplitudes \cite{Stieberger:2022lss,Stieberger:2023nol,Bhardwaj:2023vvm}.

Although we have concentrated on one-loop BCJ numerators, the foundational principles of our technique hold promise for an extension to higher-loop levels, with various CHY integrands already suggested via ambitwistor strings \cite{Geyer:2016wjx,Geyer:2018xwu,Geyer:2021oox}, traditional strings \cite{DHoker:2020prr,DHoker:2020tcq,DHoker:2021kks}, or double or multiple forward limits \cite{Geyer:2019hnn,Feng:2016nrf}. Exploring these possibilities is a direction for our upcoming research.

\vspace{.4cm}

\begin{acknowledgments}
We especially thank Song He for suggesting this study and for his collaboration on  related projects. Our thanks also go to   Freddy Cachazo, Alex Edison, Bo Feng,  Zhengwen Liu, Oliver Schlotterer, and Fei Teng for their insightful discussions and constructive feedback on our manuscript. Additionally, we are grateful to Nima Arkani-Hamed, Qu Cao, Carolina Figueiredo, Xuhang Jiang, Jiahao Liu, Yichao Tang, and Canxin Shi for their engaging dialogue and shared insights on related topics.
The research of Y.Z.\ was supported in part by a grant from the Gluskin Sheff/Onex Freeman Dyson Chair in Theoretical Physics and by Perimeter Institute. Research at Perimeter Institute is supported in part by the Government of Canada through the Department of Innovation, Science and Economic Development Canada and by the Province of Ontario through the Ministry of Colleges and Universities.

\end{acknowledgments}

 \bibliographystyle{apsrev4-1.bst}


\bibliography{Refs.bib}

\appendix

\section{More on the Proof of the Main Proposal\label{appendixa}} 

This appendix delves deeper into the substantiation of our main proposal from \cref{sec3}, detailing the intricate signs, Jacobi identities,  tadpole contributions and more examples.

\subsection{Deriving the Numerators \texorpdfstring{\eqref{ngexpre}}{16} from CHY Integral}

We show more details on the proof of \eqref{proposal2} and \eqref{proposal} using numerators from \eqref{ngexpre}, drawing on the CHY integrals \eqref{twogeneralsymcolor2}.
Summing over all orderings $\rho, \sigma \in S_{n-1}$ in \eqref{twogeneralsymcolor2}, the contribution to any graph $g \in W_1$ is represented by
\ba 
H_g=\label{ngexpre22}
 {\rm sgn}^{\rho}_{\sigma} \sum_{ (1,\rho)\in T^{-1}(g)} 
 \sum_{ (1,\sigma)\in T^{-1}(g)} 
N^{A(g,\rho)}_{1,\rho}N'{}^{A(g,\sigma)}_{1,\sigma} \,,
\ea 
where we have introduced the shorthand $N^{A(g,\rho)}_{1,\rho}\equiv  {\pmb N}_{1,\rho}\big|_{ {\pmb \ell}_1\to \ell_ {A(g,\rho)}} $. 
Inspired by the study of signs in the tree level CHY integrals for biadjoint $\phi^3$ theory \cite{He:2021lro} (see also \cite{Cachazo:2013iea,Mafra:2016ltu,Gao:2017dek}), we can write the overall sign as 
\ba 
 {\rm sgn}^{\rho}_{\sigma} = \left(\prod_{i=1}^m  {\rm sgn}^{\rho({\cal A}_i)}_{{\cal A}_i}\right)
 \left(\prod_{i=1}^m  {\rm sgn}^{\sigma({\cal A}_i)}_{{\cal A}_i}\right)\,,
\ea 
where $\rho({\cal A}_i)$ denotes a sequence in the cyclic ordering $(1,\rho)$  consisting of  the particle labels in the dangling tree ${\cal A}_i$.   
The sign of an ordering  sequence $\rho({\cal A})$ and a dangling tree ${\cal A}= [{\cal B}_{1},{\cal B}_{2}]$ compatible with that ordering is recursively defined, ultimately equating to unity for a single leg,
\ba 
\label{defsignsign}
 {\rm sgn}^{\rho([{\cal B}_{1},{\cal B}_{2}])}_{[{\cal B}_{1},{\cal B}_{2}]} \equiv
 - {\rm sgn}^{\rho([{\cal B}_{2},{\cal B}_{1}])}_{[{\cal B}_{2},{\cal B}_{1}]}
  \equiv
 {\rm sgn}^{\rho({\cal B}_{1})}_{{\cal B}_{1}}   {\rm sgn}^{\rho({\cal B}_{2})}_{{\cal B}_{2}}\,.
\ea 
The sign also fulfills
\ba 
\label{signsign2}
 {\rm sgn}^{\rho({\cal A}_i)}_{{\cal A}_i}  {\rm sgn}^{\rho'({\cal A}_i)}_{{\cal A}_i} =  {\rm sgn}^{\rho({\cal A}_i) }_{\rho'({\cal A}_i)}, \, {\rm with} \,  (1,\rho),(1,\rho')\!\in\! T^{-1}(g).
\ea 
Consequently, $H_g$ factors into
\ba 
\label{ngexpre222222}
H_g=&
\left( \sum_{ (1,\rho)\in T^{-1}(g)} 
N^{A(g,\rho)}_{1,\rho}  \prod_{i=1}^m {\rm sgn}^{\rho({\cal A}_i)}
  _{{\cal A}_i} 
\right)
\\ 
\non
&\times
\left( \sum_{ (1,\sigma)\in T^{-1}(g)} 
N'{}^{A(g,\sigma)}_{1,\sigma} 
\prod_{i=1}^m
{\rm sgn}^{\sigma({\cal A}_i)}_{{\cal A}_i}
\right)
\,,
\ea 
where choosing a reference ordering $(1,\rho_0) \in T^{-1}(g)$ and applying \eqref{signsign2} yields 
\ba 
\label{ngexpre222222nnn}
H_g=&
\left(
{\rm sgn}^{\rho_0}_g
\sum_{ (1,\rho)\in T^{-1}(g)} 
  {\rm sgn}^{\rho}
  _{\rho_0} N^{A(g,\rho)}_{1,\rho} 
\right)
\\ 
\non
&\times
\left( 
{\rm sgn}^{\rho_0}_g
\sum_{ (1,\sigma)\in T^{-1}(g)} {\rm sgn}^\sigma_{\rho_0}
N'{}^{A(g,\sigma)}_{1,\sigma} 
\right)
\,,
\ea  
with 
\ba 
\label{defsignsign222}
{\rm sgn}^{\rho_0}_g = \prod_{i=1}^m
{\rm sgn}^{\rho_0({\cal A}_i)}_{{\cal A}_i}.
\ea 
This gives an explicit definition of the sign used in \eqref{ngexpre} and \eqref{cgexpre2}. 
 The square of this sign can be dropped but we assign it to each of two parentheses in \eqref{ngexpre222222nnn}.   This formulation leads us directly to $H_g= N_g N^\prime_g$, thus affirming \eqref{proposal2} and \eqref{proposal}.

\subsection{Proof of the Jacobi Identities}

This subsection establishes that  the numerators \eqref{ngexpre},  satisfy all Jacobi identities \eqref{numjac}. Remind that the omitted dangling tree sequences in \eqref{numjac} are identical. We will denote the three graphs as ${\tilde g}_{[{\cal B}_1,{\cal B}_2]}$, ${\tilde g}_{{\cal B}_1,{\cal B}_2}$, and ${\tilde g}_{{\cal B}_2,{\cal B}_1}$.

{\bf Case I} ~
First, consider $1\notin {\cal B}_1 \sqcup {\cal B}_2$. Here, ${\tilde g}_{{\cal B}_1,{\cal B}_2}$ is shorthand for ${g}_{{\cal A}_1,\cdots,{\cal B}_1,{\cal B}_2,\cdots}$, applicable to the other two graphs. If ${\cal A}_1 \ni 1$ is absent, we can adjust loop momenta $\ell$ in \eqref{numjac} until new ${\cal A}_1 \ni 1$, aligning with definition \eqref{defggraph}.

Note that ${\tilde g}_{{\cal B}_1,{\cal B}_2}$ and ${\tilde g}_{{\cal B}_2,{\cal B}_1}$ have three or more loop propagators such that ${\tilde g}_{[{\cal B}_1,{\cal B}_2]}$ is not a tadpole. Their compatible orderings satisfy
\ba 
\label{rhorhounion}
T^{-1}({\tilde g}_{[{\cal B}_1,{\cal B}_2]}) = T^{-1}({\tilde g}_{{\cal B}_1,{\cal B}_2}) \sqcup T^{-1}({\tilde g}_{{\cal B}_2, {\cal B}_1})\,.
\ea 
Using \eqref{ngexpre} or  the equivalent $N_g$ in \eqref{ngexpre222222}, we express the numerators  of the three graphs via
\ba 
\label{ngexpre222222new}
N_g = \sum_{ (1,\rho)\in T^{-1}(g)}
\underbrace{
N^{A(g,\rho)}_{1,\rho}  \prod_{i=1}^m {\rm sgn}^{\rho({\cal A}_i)}
  _{{\cal A}_i} } _{h^\rho_g},\quad {\rm with}~ g\in W_1.
\ea 
We use $h^\rho_g$ to indicate the contribution from ordering $(1,\rho)$ to graph $g$. It is evident that
\ba 
&h^\rho_{{\tilde g}_{[{\cal B}_1,{\cal B}_2]} }= h^\rho_{{\tilde g}_{{\cal B}_1,{\cal B}_2} }, \,h^\rho_{{\tilde g}_{{\cal B}_2,{\cal B}_1} } =0,\,\,\, \forall (1,\rho)\!\in\! T^{-1}({\tilde g}_{[{\cal B}_1,{\cal B}_2]})\,,
\non
\\
&h^\rho_{{\tilde g}_{[{\cal B}_1,{\cal B}_2]} }=- h^\rho_{{\tilde g}_{{\cal B}_2,{\cal B}_1} }, \,h^\rho_{{\tilde g}_{{\cal B}_1,{\cal B}_2} } =0,\,\,\, \forall (1,\rho)\!\in\!  T^{-1}({\tilde g}_{[{\cal B}_2,{\cal B}_1]})\,.
\ea 
This confirms  the equivalence of \eqref{numjac} when $1\notin {\cal B}_1 \sqcup {\cal B}_2$.

{\bf Case II} ~
Now, suppose $1\in {\cal B}_1$. We again adjust $\ell$ in \eqref{numjac} until the three graphs become
\ba 
& {\tilde g}_{[{\cal B}_1,{\cal B}_2]} \!=\! { g}_{[{\cal B}_1,{\cal B}_2],\cdots},\,\quad 
{\tilde g}_{{\cal B}_1,{\cal B}_2} \!=\! { g}_{{\cal B}_1,{\cal B}_2,\cdots},\,
\\ \non & 
{\tilde g}_{{\cal B}_2,{\cal B}_1} \!=\! { g}_{{\cal B}_2,{\cal B}_1,\cdots},
\ea 
with identical omitted dangling tree sequences.

We use \eqref{ngexpre222222new} to directly express the numerators of ${\tilde g}_{{\cal B}_1,{\cal B}_2}$ and ${\tilde g}_{[{\cal B}_1,{\cal B}_2]}$, and relate the numerator of $ {  g}_{{\cal B}_1, \cdots,{\cal B}_2} $ to $N_{{\tilde g}_{{\cal B}_2,{\cal B}_1}}$ through
\ba 
\label{b1b2b1b2}
N_{{\tilde g}_{{\cal B}_2,{\cal B}_1}} = N_{ {{\cal B}_2,{\cal B}_1},\cdots} = { N}_{{\cal B}_1, \cdots,{\cal B}_2} \big|_{\ell\to \ell_{{\cal B}_2}}\,.
\ea 
An analogous relationship to \eqref{rhorhounion} is now observed for ${\tilde g}_{{\cal B}_1,{\cal B}_2}$, ${\tilde g}_{[{\cal B}_1,{\cal B}_2]}$, and ${ g}_{{\cal B}_1,\cdots,{\cal B}_2}$.

For $(1,\rho)\in T^{-1}({\tilde g}_{{\cal B}_1,{\cal B}_2}) \subset T^{-1}({\tilde g}_{[{\cal B}_1,{\cal B}_2]})$,  identical shift factors and signs occur in both ${\tilde g}_{{\cal B}_1, {\cal B}_2}$ and ${\tilde g}_{[{\cal B}_1,{\cal B}_2]}$, leading to 
\ba 
h^\rho_{{\tilde g}_{[{\cal B}_1,{\cal B}_2]} }= h^\rho_{{\tilde g}_{{\cal B}_1,{\cal B}_2} }, \,h^\rho_{{g}_{{\cal B}_1,\cdots, {\cal B}_2} } =0\,.
\ea

 For $(1,\rho)\in T^{-1}({  g}_{{\cal B}_1, \cdots,{\cal B}_2}) \subset T^{-1}({\tilde g}_{[{\cal B}_1,{\cal B}_2]})$, the shift factors differ between ${\tilde g}_{{\cal B}_1,\cdots, {\cal B}_2}$ and ${ g}_{[{\cal B}_1,{\cal B}_2]}$. However, they satisfy 
 \ba 
  k_{A({\tilde g}_{[{\cal B}_1,{\cal B}_2]},\rho)} =  k_{A({ g}_{{\cal B}_1,\cdots,{\cal B}_2}, \rho)} + k_{ {\cal B}_2} \,.
 \ea 
Consequently, we derive
\ba 
h^\rho_{{\tilde g}_{[{\cal B}_1,{\cal B}_2]} }=-h^\rho_{{ g}_{{\cal B}_1,\cdots, {\cal B}_2 }} \big|_{\ell\to \ell_{{\cal B}_2}}, \quad  \,h^\rho_{{\tilde g}_{{\cal B}_1,{\cal B}_2} }=0\,.
\ea
This confirms  the consistency of  numerators  with the Jacobi identities in the $1\in {\cal B}_1$ case, completing the proof.

\subsection{Double Copies in Tadpoles}

In the manuscript, we omitted tadpole contributions in \eqref{proposal} and \eqref{proposal2}, presuming their loop integrals as negligible. However, this subsection elaborates on these contributions, potentially useful for highlighting specific amplitude characteristics at the integrand level. We also show that the double copy principle extends to tadpoles, allowing a non-vanishing numerator copy to replace a vanishing color factor.

Initially, we  complete  tadpole elements in \eqref{twogeneralsym} derived in \cite{Feng:2022wee}, detailed by
\ba 
\label{twogeneralsymtadpole}
 {\rm tadpoles~of~}  & \eqref{twogeneralsym} = \sum_{g([{\cal A}_1,{\cal A}_2]) \in {\mathring T}(1,\rho) \cap {\mathring T}(1,\sigma)  } \frac{1}{P_g}
\\  \non& 
\times
\left( \ell^{\mu_1,\mu_2,\ldots,\mu_r}_{A(g({\cal A}_1,{\cal A}_2),\rho)} -\ell^{\mu_1,\mu_2,\ldots,\mu_r}_{A(g({\cal A}_1,{\cal A}_2),\rho)} \big|_{\ell \to \ell_{{\cal A}_2}}
\right)
\\ \non& \times 
\left( \ell^{\nu_1,\nu_2,\ldots,\nu_t}_{A(g({\cal A}_1,{\cal A}_2),\sigma)} -\ell^{\nu_1,\nu_2,\ldots,\nu_t}_{A(g({\cal A}_1,{\cal A}_2),\sigma)} \big|_{\ell \to \ell_{{\cal A}_2}}
\right)\,.
\ea 
Here, the specific set of tadpoles, ${\mathring T}(1,\rho)$, encompasses all $g([{\cal A}_1,{\cal A}_2])$ with ${\cal A}_1 \ni 1$ and $g({\cal A}_1,{\cal A}_2) \in T(1,\rho)$.

Based on this, we further deduce the tadpoles in \eqref{twogeneralsymcolor2}, depicted by
\ba 
\label{twogeneralsymcolor2tadpoles}
 & {\rm tadpoles~of~}   \eqref{twogeneralsymcolor2} =
 \sum_{g([{\cal A}_1,{\cal A}_2]) \in {\mathring T}(1,\rho) \cap {\mathring T}(1,\sigma)  } \frac{1}{P_{g([{\cal A}_1,{\cal A}_2])}}
\non 
\\  \non& 
\times
\left( {\pmb N}_{1,\rho}\big|_{{\pmb \ell}_1\to \ell_{A(g({\cal A}_1,{\cal A}_2),\rho)}}-{\pmb N}_{1,\rho}\big|_{{\pmb \ell}_1\to \ell_{A(g({\cal A}_1,{\cal A}_2),\rho)}+ k_{{\cal A}_2}}
\right)
\\ & \times 
\left( {\pmb N}'_{1,\sigma}\big|_{{\pmb \ell}_1\to \ell_{A(g({\cal A}_1,{\cal A}_2),\sigma)}}-{\pmb N}'_{1,\sigma}\big|_{{\pmb \ell}_1\to \ell_{A(g({\cal A}_1,{\cal A}_2),\sigma)}+ k_{{\cal A}_2}}
\right)\,.
\ea 
Summing over all permutations $\rho, \sigma \in S_{n-1}$ in \eqref{twogeneralsymcolor2tadpoles}, we obtain the tadpole terms in \eqref{proposal2} as a  summation over all tadpole graphs, symbolized as ${\mathring W}$,
\ba 
 & {\rm tadpoles~of~}   \eqref{proposal2} = \sum_{g \in {\mathring W}} \frac{1}{P_g} N_g N'_g \,,
\\  \non& 
\ea 
 where  
 \ba 
\label{twogeneralsymcolor2tadpole}
& N_{g([{\cal A}_1,{\cal A}_2]) }
= {\rm sgn}^{\rho_0}_{g({\cal A}_1,{\cal A}_2)} 
\!\!\!\!
\!\!\!\!
\sum_{ (1,\rho)\in T^{-1}(g({\cal A}_1,{\cal A}_2))}
\!\!\!\!
\!\!\!\!
{\rm sgn}^{\rho_0}_{\rho}  \,
\\
\non
&
\times \left(
{\pmb N}_{1,\rho}\big|_{ {\pmb \ell}_1\to \ell_ {A(g({\cal A}_1,{\cal A}_2),\rho)}} 
\!\! - {\pmb N}_{1,\rho}\big|_{ {\pmb \ell}_1\to \ell_ {A(g({\cal A}_1,{\cal A}_2),\rho)} + k_{{\cal A}_2}} 
\right)\,.
\ea 
and $N'_{g([{\cal A}_1,{\cal A}_2])}= N_{g([{\cal A}_1,{\cal A}_2])}\big|_{{\pmb N}\to {\pmb N}'} $. 

Subsequently, we establish that the three graphs including one tadpole $g([{\cal A}_1,{\cal A}_2])$ and two bubbles $g({\cal A}_1,{\cal A}_2)$, $g({\cal A}_2,{\cal A}_1)$ satisfy the Jacobi identities,
\ba 
 N_{g([{\cal A}_1,{\cal A}_2]) } & = N_{g({\cal A}_1,{\cal A}_2) } -N_{g({\cal A}_2,{\cal A}_1) }
\\  \non & 
= N_{g({\cal A}_1,{\cal A}_2) } -N_{g({\cal A}_1,{\cal A}_2) }\big|_{\ell \to \ell_{{\cal A}_2}}\,.
\ea 

Similarly, considering ${\mathfrak C}_n$ as a special case of $I'_n$, we can extrapolate the tadpole terms in the CHY integral of $I_n$ and ${\mathfrak C}_n$ in \eqref{proposal}. However, the color factor for a tadpole vanishes,
 \ba 
\label{twogeneralsymcolor2tadpolecolor}
 C_{g([{\cal A}_1,{\cal A}_2]) }
= {\rm sgn}^{\rho_0}_{g({\cal A}_1,{\cal A}_2)} 
\!\!\!\!
\!\!\!\!
&
\sum_{ (1,\sigma)\in T^{-1}(g({\cal A}_1,{\cal A}_2))}
\!\!\!\!
\!\!\!\!
{\rm sgn}^{\rho_0}_{\sigma}  \,
\\
\non
&\qquad\quad
\times \left(
{C}_{1,\sigma}- {C}_{1,\sigma}
\right) = 0\,.
\ea 
Despite this, it is pertinent to note that the double copy concept remains valid even for tadpoles. One can substitute the non-existent color factor $C_{g([{\cal A}_1,{\cal A}_2])}$ with a valid numerator copy $N'_{g([{\cal A}_1,{\cal A}_2])}$, thereby procuring the tadpole contributions at the integrand level for the theory obtained by double copy.

Tadpole contributions in \eqref{proposal22}
 are similarly retrievable, with 
$ {\bar N}_{g([{\cal A}_1,{\cal A}_2]) } =  {\bar N}_{g({\cal A}_1,{\cal A}_2) } -{\bar N}_{g({\cal A}_1,{\cal A}_2) }\big|_{\ell \to \ell_{{\cal A}_2}}$. The corresponding symmetry factor for a tadpole ${\cal S}_{g([{\cal A}_1,{\cal A}_2])}$ is 1.

In the context of massless bubbles, the literature presents various methodologies for regularizing singular poles like $s_{2\cdots n}$. A notable example is the Minahaning procedure \cite{Minahan:1987ha,Berg:2016wux,Berg:2016fui}, which effectively yields a finite contribution at the loop integrand level. Similarly, it is theoretically feasible to regularize even more singular poles $s_{12\cdots n}$ found in tadpole graphs. This can be achieved by judiciously applying the forward limit in higher dimensions, thereby attaining a finite contribution at the integrand level. However, detailed explorations of this method is earmarked for future research. It is important to reiterate that all massless bubbles and tadpoles, being scaleless integrals, can ultimately be disregarded at the amplitude level.

\subsection{Examples}

Here we provide a concrete example of \eqref{proposal2} at $n=3$, incorporating tadpole contributions for comparison with \eqref{combiIn3ptc},
\allowdisplaybreaks[1]
\ba
 \label{combiInJn3pt222}
 \non
\frac{1}{\ell^2}& \int {\rm d}\mu_3 \, I_3 \,  I'_3
\cong
\frac{(N_{1,2,3} -N_{1,3,2}  ) (N'_{1,2,3}-N'_{1,3,2})}{P_{1,[2,3]}}
  \\
&
+\frac{N_{1,2,3} N'_{1,2,3}}{P_{1,2,3}} +    \frac{(N_{1,2,3} -N_{1,3,2}  ^2) (N'_{1,2,3}-N'{}^2_{1,3,2})}{P_{[1,2],3}}
\nl & 
+\frac{N_{1,3,2} N'_{1,3,2}}{P_{1,3,2}} +    \frac{(N_{1,2,3} ^3-N_{1,3,2}  ) (N'{}^3_{1,2,3}-N'_{1,3,2})}{P_{[3,1],2}}
\nl
&+
\frac{
(N_{1,2,3} -N_{1,2,3}^{23}- N_{1,3,2}+N_{1,3,2}^{23}  )\times(N\to N') 
}{P_{[1,[2,3]]}}
\nl
&+
\frac{
 (N_{1,2,3} -N{}^{3}_{1,2,3}- N{}^2_{1,3,2}+N{}^{23}_{1,3,2}  )
\times(N\to N') 
}{P_{[[1,2],3]}}
\nl
&+
\frac{
 (N{}^{23}_{1,2,3} -N{}^{3}_{1,2,3}- N{}^2_{1,3,2}+N{}^{}_{1,3,2}  ) 
\times(N\to N') 
}{P_{[[3,1],2]}}
\,,
\ea
\allowdisplaybreaks[0]
\!\!where  $ N{}^{3}_{1,2,3} = { N}_{1,2,3}\big|_{\ell \to \ell_{3}}$. 

We also spell out a specific instance of \eqref{proposal2} at $n=4$, providing a detailed exposition of all the shift factors $A(g,\rho)$ across the numerators,
\allowdisplaybreaks[1]
\ba
\label{combiInJn4pt}
\frac{1}{\ell^2}&\!\int \! {\rm d}\mu_4 \, I_4 \,I'_4
\cong \!
\Big[ \frac{N_{1,2,3,4} N'_{1,2,3,4}}{P_{1,2,3,4}}
 \!+\! {\rm perm}(2,3,4) \Big] 
\\
&
+ \Big[
\Big( 
\frac{
N_{1,[2,3],4} N'_{1,[2,3],4}
}{P_{1,[2,3],4}} 
+\frac{
N_{1,4,[2,3]} N'_{1,4,[2,3]}
}{P_{1,4,[2,3]}} 
\nl
& \qquad  +
\frac{N_{1,[[2,3],4]} N'_{1,[[2,3],4]}
}{P_{1,[[2,3],4]}} 
 + {\rm cyc}(2,3,4) \Big] 
\nl&
+ 
\Big[
\frac{
(N_{1,2,3,4}- N_{1,3,4,2}^2 ) (N\to N' ) 
}{P_{[1,2],3,4}} 
+{\rm perm}(2,3,4)\Big]
\nl&
+ 
\Big[
\frac{
(N_{1,2,[3,4]}- N_{1,[3,4],2}^2 ) (N\to N' ) 
}{P_{[1,2],[3,4]}} 
+{\rm cyc}(2,3,4)\Big]
\nl&
+ 
\Big[
\frac{
(N_{1,[2,3],4}- N_{1,4,[2,3]}^{23} ) (N\to N' )
}{P_{[1,[2,3]],4}} 
+{\rm cyc}(2,3,4)\Big]
\nl&
+ 
\Big[
\frac{
(N_{1,2,3,4}\!-\!N^{2}_{1,3,4,2}\!-\!N^{3}_{1,2,4,3} \!+\!N^{23}_{1,4,3,2}) (N\to N')
}{P_{[[1,2],3],4}} 
\nl &
\qquad +{\rm perm}(2,3,4)\Big]
+{\rm tadpoles}
\,,
\non
\ea
\allowdisplaybreaks[0]
\!\!where  $ N{}^{2}_{1,[3,4],2} = N{}^{2}_{1,3,4,2}-N{}^{2}_{1,4,3,2}=N{}^{}_{1,3,4,2} \big|_{\ell \to \ell_{2}}-N{}^{}_{1,4,3,2} \big|_{\ell \to \ell_{2}}$.

\section{ Proof of Equivalence between \texorpdfstring{${\bar I}_n$}{barIn} and \texorpdfstring{$I_n$}{In} \label{appendixb} }

To establish the equivalence of ${\bar I}_n$ and $I_n$, we must demonstrate their identical contributions to graph pairs $g({\cal A}_1,{\cal A}_2, ..., {\cal A}_m)$ and $g({\cal A}_1,{\cal A}_m, ..., {\cal A}_2)$ for $3\leq m \leq n$, or single graphs at $m=2$, via \eqref{proposal}. We denote these uniformly as pair $(g,g^T)$. The proof seeks to confirm
\ba 
\label{toproveeq}
\frac{N_g C_g}{P_g} + \frac{N_{g^T} C_{g^T}}{P_{g^T}} \cong \frac{{\bar N}_g C_g}{P_g} + \frac{{\bar N}_{g^T} C_{g^T}}{P_{g^T}} \,.
\ea 

Utilizing \eqref{ngexpre}, we find
\begin{align} 
\label{ngexpreg}
& N_g = {\rm sgn}^{\rho_0}_g \sum_{(1,\rho)\in T^{-1}(g)} {\rm sgn}^{\rho_0}_{\rho} N_{1,\rho}\big|_{\ell\to \ell_{A(g,\rho)}},
\\
\non
& N_{g^T} = {\rm sgn}^{\rho_0^T}_{g^T} \sum_{(1,\rho)\in T^{-1}(g)} (-1)^n {\rm sgn}^{\rho_0}_{\rho} N_{1,\rho^T}\big|_{\ell \to \ell_{A(g^T,\rho^T)}}\,.
\end{align} 

Redefining loop momentum $\ell \to -\ell_{{\cal A}_1}$ in the second term of \eqref{toproveeq} does not affect the loop integral of the amplitude, yielding
\ba 
\label{tworeplace}
&\frac{N_{g^T} C_{g^T}}{P_{g^T}} \Bigg|_{\ell \to -\ell_{{\cal A}_1} } \cong 
\frac{  C_{g}{\rm sgn}^{\rho_0}_g}{P_{g}} 
\\ 
\nonumber
&\qquad \times
 \sum_{ (1,\rho)\in T^{-1}(g)} 
(-1)^n  {\rm sgn}^{\rho_0}_{\rho} 
 \, { N}_{1,\rho^T}\big|_{ { \ell} \to \ell_ {A(g^T,\rho^T)}} \big|_{ { \ell} \to -\ell_ {{\cal A}_1}} \,,
\ea 
using $P_{g^T}= P_{g}\big|_{\ell \to -\ell_{{\cal A}_1}}$ according to  \eqref{reverse}. 
Observe that $k_{{\cal A}_1} = k_{A(g,\rho)} + k_1 + k_{A(g^T, \rho^T)}$ and hence the sequence of two substitution rules  in \eqref{tworeplace} simplifies to $\ell\to -\ell-k_{{\cal A}_1}+ k_{A(g^T, \rho^T)}= -\ell_{1,A(g,\rho) }$. 
Thus, the contribution  of pair graphs in the LHS 
 of \eqref{toproveeq} combines as
\ba 
\label{twographsum}
&{\text {LHS of}} ~ \eqref{toproveeq} \cong 
\frac{  C_{g} {\rm sgn}^{\rho_0}_g}{P_{g}} 
 \sum_{ (1,\rho)\in T^{-1}(g)} 
\!\!\!\!\!\!
 {\rm sgn}^{\rho_0}_{\rho}
 \Big( { N}_{1,\rho} \big|_{ { \ell} \to \ell_{ A(g,\rho) } } 
  \non
  \\ 
 & \qquad \qquad +
(-1)^n  
 \, N_{1,\rho^T} \big|_{ { \ell} \to -\ell_{1,A(g,\rho) } } 
 \Big)\,.
\ea 

Similarly, each of the two terms on the RHS of  \eqref{toproveeq} can  be shown to equal half of \eqref{twographsum}, proving ${\bar I}_n\cong I_n$ and thereby affirming the refined double copy \eqref{proposal22}.

\section{Crossing Symmetric BCJ Numerators \label{sec5}}

\subsection{Proposal}

Considering $n$ identical external bosons in scattering, we examine the effects of averaging all particle label permutations in half-integrand ${\bar I}_n$ defined in \eqref{defInbar},
\begin{align}
&\label{firstrep}
I_n \cong {\tilde I}_n = \frac{1}{n!}\sum_{\gamma \in S_{n}} {\bar I}_n\big|_{\gamma(1)\to1,\gamma(2)\to 2,\ldots, \gamma(n)\to n}
\end{align}
and rearrange it as
\begin{align}
\label{secondrep}
\tilde I_n = \sum_{\rho \in S_{n-1}}{\tilde N}_{1,\rho} {\rm PT}(1,\rho).
\end{align}
%
We claim that
 all operators ${\tilde N}_{1,\rho}$ are relabeled versions of ${\tilde N}_{1,2,\ldots,n}$, and
\begin{align}
&
\label{crossingsymm2ccc}
{\tilde N}_{1,2,\ldots,n}{=} \frac{1}{n}\sum_{j=1}^{n} \! \left({\hat N}_{1,2,\ldots,n}\big|_{i\to i+j}\right)\big|_{{\pmb \ell}_{j+1}\to {\pmb \ell}_1+k_{12\ldots j}},\\
\nonumber
&{\rm with}\,
{\hat N}_{1,2,\ldots,n} {=} \frac{1}{(n-1)!}\sum_{\gamma \in S_{n-1}} {\bar { N}}_{1,\gamma}\big|_{\gamma(2)\to 2,\ldots, \gamma(n)\to n}.
\end{align}
Moreover, its master BCJ numerators, ${\tilde {N}}_{1,\rho} = {\tilde N}_{1,\rho}\big|_{{\pmb \ell}_1\to \ell}$, exhibit $S_n$ relabeling relations, along with properties such as those in \eqref{shiftell} and \eqref{requirments2}, thus achieving crossing symmetry.

Although ${\tilde I}_n$ is symmetric in particle labels, its representation in \eqref{secondrep} appears asymmetric as particle 1 seems distinct, useful in evaluating its CHY integral \eqref{proposal} based on \eqref{twogeneralsym}. 
However, the resulting loop integrand retains $S_n$ symmetry, ensuring crossing symmetric BCJ numerators.

\subsection{Proof \label{appendixc} }

The symmetrization process outlined in \eqref{firstrep} is twofold,
\begin{align}
& I_n \cong {\hat I}_n = \frac{1}{(n-1)!}\sum_{\gamma \in S_{n-1}} {\bar I}_n\big|_{\gamma(2)\to 2,\ldots, \gamma(n)\to n}\,,
\\
\label{secondstep}
& I_n \cong {\tilde I}_n =\frac{1}{n} \sum_{j=1}^n {\hat I}_n \big|_{i\to i+j }\,,
\end{align}
with particle relabels understood modulo $n$.

Reconfiguring ${\hat I}_n$ yields,
\begin{align} 
& I_n \cong {\hat I}_n = \sum_{\rho \in S_{n-1}}\hat {\pmb{N}}_{1,\rho} {\rm PT}(1,\rho),
\\
\nonumber
& \text{where}~ \hat{\pmb{N}}_{1,\rho} =  
\frac{1}{(n-1)!}\sum_{\gamma\in S_{n-1}} {\bar {\pmb N}}_{1,\gamma} \big|_{\gamma(2)\to \rho(2),\cdots, \gamma(n)\to \rho(n) }.
\end{align}
Evidently, all $\hat {\pmb{N}}_{1,\rho}$ are mere relabellings of $ \hat{\pmb{N}}_{1,2,\cdots,n}$.

Acknowledging  the symmetry of ${\hat I}_n$ among labels $2,3,$ $\cdots,n$, the summation in \eqref{secondstep} restructures as
\begin{align} 
 {\tilde I}_n=& \sum_{\rho\in S_{n-1}} \frac{1}{n} \sum_{j=1}^n \hat{\pmb{N}}_{\rho(j+1),\cdots,\rho(n),1,\rho(2),\cdots,\rho(j)}
\\ \nonumber 
& \qquad\times{\rm PT}(\rho(j+1),\cdots,\rho(n),1,\rho(2),\cdots,\rho(j)),
\end{align} 
where ${\hat {\pmb  N}}_{\rho(j+1),\cdots,\rho(j)}$ now becomes a polynomial of operators ${\pmb \ell}_1\big|_{1\to \rho(j+1)}= {\pmb \ell}_{\rho(j+1)}$.

Utilizing identities from \cite{Feng:2022wee},
\ba
&\left( \prod_{q=1}^r {\pmb \ell}_{j+1}^{\mu_q}
\right)
{\rm PT}(j+1, \cdots,n,1,2,\cdots,j)
\\ 
\non
&\qquad\quad =\left( \prod_{q=1}^r 
\left( {\pmb \ell}_1^{\mu_q}+ k_{12\cdots j}^{\mu_q} \right)
\right)
{\rm PT}(1,2, \cdots,n)\,,
\ea 
and their relabelling,
we can transform $\tilde I_n$ as 
\begin{align}
\tilde I_n = \sum_{\rho \in S_{n-1}}\tilde{\pmb{N}}_{1,\rho} {\rm PT}(1,\rho)\,,
\end{align}
with 
\begin{align} 
\label{ntildetilde}
\tilde{\pmb{N}}_{1,\rho} \!= \!\frac{1}{n}\sum_{j=1}^n \hat{\pmb{N}}_{\rho(j+1),\cdots,\rho(n),1,\rho(2),\cdots,\rho(j)}\big|_{{\pmb \ell}_1 \to {\pmb \ell}_1+k_{1\rho(2)\cdots \rho(j)} }.
\end{align} 
Distinctly, all ${\tilde  {\pmb N}}_{1,\rho}$ are relabellings of $ {\tilde {\pmb N}}_{1,2,\cdots,n}$. Moreover, the cyclic order summation in \eqref{ntildetilde} suggests cyclic relabelling relations for the master BCJ numerators,
\begin{align} 
\label{cyclicrel}
{\tilde N}_{\pi(j)\pi(j+1) \cdots \pi(j-1)} =  
{\tilde N}_{\pi(1)\pi(2) \cdots \pi(n)} \big|_{i\to i+j-1}.
\end{align} 
For instance, 
\ba 
&{\tilde N}_{2\cdots n1} \equiv
{\tilde N}_{12\cdots n}\big|_{\ell \to \ell-k_1}=\frac{1}{n} \sum_{j=1}^n {\hat N}_{j+1,j+2,\cdots,j}\big|_{\ell\to \ell_{23\cdots j}}
\non 
\\ 
& 
\qquad\qquad\qquad = {\tilde N}_{12\cdots n}\big|_{1\to2,2\to 3,\cdots,n\to 1}\,.
\ea 
The general proof for \eqref{cyclicrel} is direct, and combined with the $S_{n-1}$ relabelling relations in $2,\cdots,n$, it is clear that ${\tilde N}_{1,\rho}$ maintains $S_n$ relabelling relations,
\begin{align} 
{\tilde N}_{\pi(1) \pi(2)\cdots \pi(n)} =  {\tilde N}_{12\cdots n}\big|_{1\to \pi(1),2\to \pi(2),\cdots,n\to \pi(n)}.
\end{align} 
As ${\hat N}_{1,\sigma}$ and subsequently ${\tilde N}_{1,\sigma}$ are linearly composed of ${\bar n}_{1,\gamma}$, they naturally inherit properties like \eqref{requirments2}. With the definition \eqref{shiftell} in play, we affirm the crossing symmetry of master BCJ numerators ${\tilde N}_{1,\sigma}$.

It is noteworthy that ${\bar I}_n$, ${\hat I}_n$, or ${\tilde I}_n$ could each substitute $I_n$ in \eqref{proposal}, yielding BCJ numerators of varying properties. While one could derive these from the original $(n-1)!$ distinct BCJ numerators ${n}_{1,\rho}$ by direct manipulation of the loop integrand ${\cal M}_n^{\cal O}(\ell)$ in \eqref{proposal}, our approach leveraging worldsheet functions proves more elegant.

Our methods also suggest potential applicability from CHY to string integrals. Despite the complications in acquiring crossing-symmetric BCJ numerators in string amplitudes due to the BRST gauge choice  \cite{Bridges:2021ebs,Lee:2015upy}, our results indicate that this is achievable with certain string integrand modifications.

\vspace{.2cm}

\section{Proof of Expansion for the YM Integrand \label{appendixd}}

This appendix establishes the equivalence between \eqref{eq:Psiexp} and \eqref{eq:psi1lf}.

Consider an arbitrary $n$-point ordering sequence denoted as
\ba 
\label{rhor}
\rho^{(a)} \!\!=\! \rho(a\!+\!1)\rho(a\!+\!2)\ldots \rho(n)\rho(1)\ldots \rho(a),\, \text{where}~ \rho(1){=}1.
\ea 
We aim to compare the coefficient of  ${\rm PT}^{\rm tree}(+\rho^{(a)}-)$ within \eqref{eq:Psiexp} and \eqref{eq:psi1lf}.
In \eqref{eq:Psiexp}, the coefficient is explicitly
\ba \label{D2coef}
\eqref{eq:Psiexp} \big|_{ {\rm PT}^{\rm tree}(+\rho^{(a)}-)} = \prod_{i=1}^{n}\epsilon_i\cdot(\ell+Y_i(\rho^{(a)} )) \,.
\ea 

For \eqref{eq:psi1lf}, all generalized PT factors 
${\pmb \ell}_1^{\mu_1,\mu_2,\ldots,\mu_r}\mathrm{PT}(  \rho^{(0)} )$ 
with 
\ba 
\label{rho0}
\rho^{(0)} = \rho(1)\rho(2)\ldots \rho(a)\rho(a+1)\ldots \rho(n),\, \text{where}~ \rho(1)=1, 
\ea 
retain non-zero coefficients for ${\rm PT}^{\rm tree}(+\rho^{(a)}-)$ according to  \eqref{eq:pttensor}, resulting in 
\ba 
\label{D4coef}
\eqref{eq:psi1lf} \big|_{ {\rm PT}^{\rm tree}(+\rho^{(a)}-)} = \prod_{i=1}^{n}\epsilon_i\cdot(\ell+Y_i(\rho^{(0)} )- k_{\rho(1)\ldots \rho(a)}) \,.
\ea 
Recall $Y_i(\pi(1),\ldots,\pi(j),i,\ldots) = k_{\pi(1),\ldots,\pi(j)}$. 
A comparison of \eqref{rhor} and \eqref{rho0} reveals
\ba 
Y_i(\rho^{(a)})= Y_i(\rho^{(0)})- k_{\rho(1)\ldots \rho(a)}, \forall i\in \{\rho(1),\ldots,\rho(n)\}\,,
\ea 
on the support of momentum conservation for the external legs.

Thus, we confirm $\eqref{D2coef}=\eqref{D4coef}$, concluding the proof.

Notice that $Y_1(\rho^{(0)})=0$ and $\epsilon_{ \rho(n) }\cdot Y_{\rho(n)}(\rho^{(0)})=- \epsilon_{ \rho(n) } \cdot k_{\rho(n)}=0$, establishing that  the rank of the generalized PT factors in \eqref{eq:psi1lf} is at least 2.

\end{document}